\documentclass[runningheads,a4paper]{llncs}

\usepackage{amsmath,amssymb,amsfonts,epsfig}
\usepackage{verbatim}
\usepackage{framed}
\usepackage{amssymb}
\usepackage{soul} % for highlighting
\usepackage{graphicx}
\usepackage{float}
\usepackage{cite} % [sort
\usepackage{epstopdf}
\usepackage{hyperref}
\usepackage{color}
\newcommand{\keywords}[1]{\par\addvspace\baselineskip
\noindent\keywordname\enspace\ignorespaces#1}

\newcommand{\be}{\begin{equation}}
\newcommand{\ee}{\end{equation}}

\newcommand{\vv}{{\mathbf v}}

\newcommand{\vx}{{\mathbf x}}
\newcommand{\vy}{{\mathbf y}}

\newcommand{\ve}{{\mathbf e}}
\newcommand{\vzero}{{\mathbf 0}}

\newcommand{\bcr}{{\textbf {MCR} }} % Name of problem. Changed 'Bacterial' to 'Microbial'
\newcommand{\pack}{{\textbf {COMPASS} }} % Name of package % STRIP
\newcommand{\rRNA}{16S rRNA }

\newcommand{\bx}{{\mathbf x}}
\newcommand{\by}{{\mathbf y}}
\newcommand{\bz}{{\mathbf z}}

\newcommand{\norm}[1]{\left\| #1 \right\|}
\newcommand{\alphabet}{\Upsilon}
\newcommand{\metric}{\mathcal{D}}
\newcommand{\loss}{l}

\newcommand{\hatx}{\hat{\bx}} % estimator of x from data
\newcommand{\haty}{\hat{\by}} % plug-in estimator of x for y from data
\newcommand{\readlen}{\ensuremath{L}}
\newcommand{\lex}{lex} % lexicographic ordering
\newcommand{\Aone}{A_{(1)}}

\newcommand{\eps}{\epsilon}

\newcommand{\nin}{\noindent}

\def\qed{\hfill \vrule height1.3ex width1.2ex depth-0.1ex}

\begin{document}

\mainmatter % start of an individual contribution

% first the title is needed

\vspace{-0.4cm}
\title{Accurate Profiling of Microbial Communities from Massively Parallel Sequencing using Convex Optimization}

% a short form should be given in case it is too long for the running head
\titlerunning{Accurate Profiling of Microbial Communities}

% the name(s) of the author(s) follow(s) next
%
% NB: Chinese authors should write their first names(s) in front of
% their surnames. This ensures that the names appear correctly in
% the running heads and the author index.
%

\author{Or Zuk$^{1,2}$%
%\thanks{Please note that the LNCS Editorial assumes that all authors have used
%the western naming convention, with given names preceding surnames. This determines
%the structure of the names in the running heads and the author index.}%
\and Amnon Amir$^3$\and Amit Zeisel$^3$\and Ohad Shamir$^4$\and Noam Shental$^5$}

\authorrunning{Accurate Identification and Profiling of Microbial Communities}
% (feature abused for this document to repeat the title also on left hand pages)

% the affiliations are given next; don't give your e-mail address
% unless you accept that it will be published

\institute{$^1$ Broad Institute of MIT and Harvard; \quad $^2$ Toyota Technological Institute at Chicago \\
$^3$ Department of Physics of Complex Systems, Weizmann Institute of Science \\
$^4$ Microsoft Research, New England \\
$^5$ Department of Computer Science, The Open University of Israel
}

%\url{http://www.springer.com/lncs}}

%
% NB: a more complex sample for affiliations and the mapping to the
% corresponding authors can be found in the file "llncs.dem"
% (search for the string "\mainmatter" where a contribution starts).
% "llncs.dem" accompanies the document class "llncs.cls".
%

\toctitle{Lecture Notes in Computer Science}
\tocauthor{Authors' Instructions}
\maketitle

\vspace{-0.25cm}
\begin{abstract}
We describe the Microbial Community Reconstruction (\bcr\!\!) Problem, which is fundamental for microbiome analysis.
In this problem, the goal is to reconstruct the identity and frequency of species comprising a
microbial community, using short sequence reads from Massively Parallel Sequencing (MPS) data obtained for specified genomic regions.
We formulate the problem mathematically as a convex optimization
problem and provide sufficient conditions for identifiability, namely the ability to reconstruct species identity and frequency
correctly when the data size (number of reads) grows to infinity.
We discuss different metrics for assessing the quality of the reconstructed solution, including a novel phylogenetically-aware
metric based on the Mahalanobis distance, and give upper-bounds on the reconstruction error for a finite number of reads under different metrics.
We propose a scalable divide-and-conquer algorithm for the problem using convex optimization, which enables us to handle large problems (with $\sim\!10^6$ species). We show using numerical simulations that for realistic scenarios, where the microbial communities are sparse, our algorithm gives solutions with high accuracy, both in terms of obtaining accurate frequency, and in terms of species phylogenetic resolution.

\vspace{-0.05cm}
 %PROOFS
\end{abstract}

\vspace{-0.9cm}
\section{Introduction}
\vspace{-0.25cm}
Characterization of the micro-organisms present in a microbial community is of major biological and clinical importance.
Since different micro-organisms have different genomes, it is possible to identify species based on their DNA sequences, using
either whole-genome sequencing, or sequencing of pre-specified regions.
The 16S ribosomal RNA gene (\rRNA\!\!) is of particular
interest for identifying microbial communities via sequencing.
It has both highly conserved regions,
present in almost all microbial species, together with
variable regions. The conserved regions allow sequence amplification
using universal PCR primers, while the variable regions provide
information used to distinguish between different species.
Large databases
\cite{desantis2006greengenes,cole2009ribosomal}
with millions of \rRNA sequences may enable species identification by querying sequencing results in a database.

Previous methods aiming to characterize microbial communities using
microarrays \cite{gentry2006microarray} and Sanger sequencing
\cite{amir2011bacterial} have shown that, in principle, it is possible
to identify species present in a sample, yet it is not clear how to get accurate estimation of species frequencies
from the analog measurements provided by these technologies.
Massively Parallel Sequencing (MPS)
\cite{mardis2008impact}, also known as Next-Generation Sequencing
(NGS), provides high-throughput digital sequence data and can allow a
more detailed and accurate picture of the species in the mixture. In
this method, one obtains a large number of short sequence reads from
the mixture, and the goal is to reconstruct the identities and
quantities of the species present. Many studies have used short reads
to characterize microbial communities \cite{hamady_microbial_2009},
yet they did not demonstrate an ability to identify the specific
species present and quantify their abundance in the mixture - reliable recognition was typically achieved only at coarse genus level \cite{huse_exploring_2008}.
The main drawback of MPS is the relatively short read length
(typically around 50-400 base-pairs in current technologies), which poses a problem
for species reconstruction; short reads do not provide unambiguous evidence in support of the presence of a specific species,
as typically the same read may originate from multiple different
species, and cannot be uniquely aligned to the reference database.

Recently, more sophisticated methods for quantifying species abundance were developed,
for \rRNA \cite{meinicke2011mixture,eskin2013ealps} and whole metagenome shot-gun sequencing data\cite{xia2011accurate}. These methods take into account read-assignment ambiguity and enable increased species resolution, but the question of maximal reconstruction resolution achieved was not systematically studied.

In this paper, we study mathematically the Microbial Community Reconstruction problem (\bcr\!\!) -
in which we use MPS data to characterize a microbial community. In a nutshell, the computational and statistical problem we face is as follows: given a large collection of short MPS reads (strings) sampled
from a known database of species' sequences (longer strings) according to a certain unknown distribution, our goal is to estimate the sampling frequencies for each species in the database, and specifically recover the support of the distribution, i.e. the list of species with non-zero sampling probabilities.
We model the sequencing process statistically, providing a probabilistic generative model for the short read data at hand. We prove conditions for {\it identifiability} - namely the ability to reconstruct precisely the identity and frequency of species present in the mixture from the short read data as the number of reads is increased.
We prove upper-bounds on reconstruction errors for a finite number of reads.
We propose a divide-and-conquer algorithm, handling large scale problems with hundreds of thousands of species,
which is particularly appealing for sparse microbial communities - that is, realistic scenarios where
only hundreds or a few thousands of species are present in the mixture, out of the possible millions of species in the database (see e.g. \cite{Eckburg:01,paster_bacterial_2001}). We study the reconstruction performance in these realistic settings by simulating reads from the Greengenes \rRNA database \cite{desantis2006greengenes}.

Our goal here is to formulate and study the problem mathematically. Practical considerations (e.g. amplification and sequencing biases, restrictions on primers, paired-end reads) together with experimental results for real sequencing data are described in a separate publication \cite{amir2013}.

In the spirit of reproducible research, we have implemented all of our algorithms in the Matlab package
\pack ({\underline C}onvex {\underline O}ptimization for {\underline
  M}icrobial
{\underline P}rofiling by {\underline A}ggregating
{\underline S}hort {\underline S}equence reads),
which is freely available at github: \quad \quad
\url{https://github.com/NoamShental/COMPASS}.

\vspace{-0.3cm}
\section{The \bcr Problem Formulation}
\begin{figure}[!ht] % [H] Schematic Description
\begin{center}
\includegraphics[totalheight=0.24\textheight,width=0.7\textwidth]{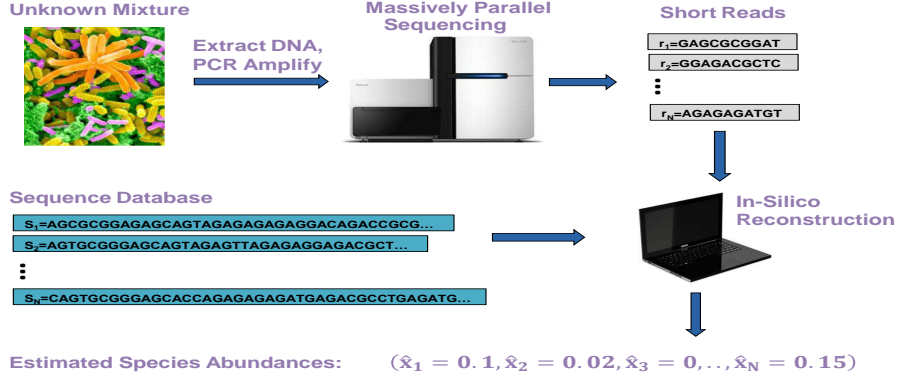} % pdf: 0.5, 1.1
\end{center}
\caption[Schematic Description of Microbial Community Reconstruction]{
The steps performed for species reconstruction using the \bcr method. First,
DNA is extracted and amplified using PCR with universal primers matching the \rRNA gene. The DNA is then sheared and sequenced using MPS, producing millions of short sequence reads.
The sequencing data (reads), together with a database of \rRNA sequences, are entered into the computational pipeline
providing estimated species abundances as output.
\label{fig:BCR_scheme}}
\end{figure}

\vspace{-0.15cm}
We describe informally and briefly the biological settings. Our goal
is to identify the species present in a given sample.
We extract DNA, use \rRNA universal primes and amplify the DNA in this
region. We then assume that DNA is sheared randomly and sequence it using MPS.
We assume that the sequences database contains \rRNA sequences for {\it all} species present in the mixture,
and reconstruct the species in the mixture {\it in silico}. A schematic representation of the \bcr method is shown in Figure \ref{fig:BCR_scheme}.  %PROOFS

We denote by $N$ the number of species in the database. The species' \rRNA sequences are marked $S_1,..,S_N$, represented as
strings over the alphabet $\alphabet = \{\text{`A', `C', `G', `T'}\}$.
We assume that the $S_i$'s are {\it distinct} sequences. The sequences may have {\it different} lengths $n_1,..,n_N$, with $n_i$ the length of the i-th species' sequence, i.e. $S_i \in \alphabet^{n_i}$. For the \rRNA gene, the lengths $n_i$ are roughly $1500$ base-pairs.
We define the maximum sequence length as $n_{MAX} \equiv \max_{i} n_i$.
We denote by $s_{i,j}$ the j-th nucleotide in the i-th species'
sequence $S_i$, and by $s_{i,j:k}$ the substring containing
nucleotides $j, j+1, .., k$ in the sequence $S_i$.

We represent the proportion of each species in the mixture using a vector $\vx$ of length $N$, with $x_i$ the frequency of species $i$.
% Vectors are denoted with boldface lowercase English letters ($\bx, \by, ..$), and matrices with uppercase letters ($A, B,..$).
We have $\vx \in \Delta_N$, where $\Delta_N$ is the $N$-dimensional simplex, $\Delta_N = \{ \vx \: : \: x_i \geq 0, \: \sum_{i=1}^N x_i = 1 \}$.
We represent the interior of a set $\mathcal{A}$ as $int(\mathcal{A})$. In particular, $int(\Delta_N)$ is the subset of $\Delta_N$
containing vectors with positive entries, $int(\Delta_N) = \{ \vx \: : \: x_i > 0, \: \sum_{i=1}^N x_i = 1 \}$.

We observe data in the form of $R$ reads of length $L$, $r_1, .., r_R \in \alphabet^L$, with \readlen\: typically around $\sim\!50\!-\!400$, as in the Illumina and 454 sequencing technologies. We represent the data by a vector of read frequencies, $\vy \in \Delta_{4^L}$, with the {\it j}-th coordinate given by $y_j = \frac{1}{R} \sum_{i=1}^R 1_{\{\lex(r_i) = j\}}, \quad \forall j=1,..,4^L$. Here $\lex(r)$ is the index of $r$ in the lexicographic ordering of all $4^L$ possible reads (i.e. $\lex(\text{'AAA ... A'}) = 1, .., \lex(\text{'TTT ... T'}) = 4^L$). We also define the inverse lexicographic ordering transformation, $\lex^{-1}$, which for a given index $j$ gives the corresponding
sequence (e.g. $\lex^{-1}(18) = \text{'AAA ... ATC'}$).  %PROOFS

\vspace{-0.01cm}
In the \bcr problem, the data vector
$\vy$ and the database sequences $S_1,..,S_N$ are given as input. Our goal is to reconstruct the species frequencies vector $\vx$ from this information. % where we seek a `good' reconstructed $\vx$ as will be discussed later.
%We denote this problem as $MCR(L, R, S)$.
The vector $\vy$ is of exponential length $(4^L)$ but very sparse, with only $M \leq R$ non-zero coordinates, where $M$ is the number of {\it unique} sequence reads. We store and manipulate only the non-zero part of $\vy$ - therefore the computational complexity of all of our algorithms will depend on $M$, and not the exponentially large $4^L$ (see Section \ref{sec:algorithm}). In typical MPS experiments with current technologies $R$ may be on the order of $\sim\!10^5 - 10^8$. %PROOFS

\vspace{-0.1cm}
\subsection{Probabilistic Model}
\vspace{-0.05cm}
We formulate a probabilistic generative model capturing the sequencing process.
We assume that the $R$ reads are sampled identically and independently (i.i.d.) from the set of amplified regions in two steps,
\begin{enumerate}
\item
First, sample a microbial species $b$ from the set of possible species $\{1,..,N\}$, with the probability of species $j$ being sampled proportional to the amount of DNA from this species,
$x'_j \equiv Pr(b = j) = \frac{x_j n_j}{\sum_{i=1}^N x_i n_i}$.

\item
Next, sample a read $r$ from a distribution given by the species $b$.
We represent sampling probabilities using a $4^L \times N$ read-sampling matrix $A = A(S, L)$ whose $(i,j)$-th
entry is the probability to observe read $i$ given that we know it came from species $j$, $A_{ij} = Pr(r = i | b = j)$.
\end{enumerate}

\begin{remark}
The vector of sampling probabilities $\vx'$ from step 1 is obtained by re-weighting the frequency
vector $\vx$ according to the sequence lengths. For ease of notation, we disregard this re-weighting,
and denote both vectors as $\vx$. When all sequences lengths $n_j$ are identical we have indeed $\vx'=\vx$.
More generally, the vectors are different but we can easily convert $\vx$ to $\vx'$
or $\vx'$ to $\vx$ using the above relation $x'_j = \frac{x_j n_j}{\sum_{i=1}^N x_i n_i}$
\end{remark}

\nin The sampling process defines a probability distribution $P_{\bx} = P_{\bx}(\vy ; A, L)$ on the space of possible frequencies $\Delta_{4^L}$, % PROOFS
\begin{align}
P_{\bx}(\vy ; A, L) = \left\{ \begin{array}{ll}
\sum_{j=1}^N A_{ij} x_j \:\:\quad & \quad\:\: \vy = \ve^{(i)} \\
0 \:\:\quad & \quad\:\: \mbox{otherwise}
\end{array} \right.
\label{eq:mixture_model}
\end{align}
\vspace{-0.15cm}
\nin where $\ve^{(i)} \in \Delta_{4^L}$ is the i-th vector in the standard basis, $e^{(i)}_i =1, e^{(i)}_j =0 \: \forall j \neq i$. %PROOFS
The data can be represented as $R$ i.i.d. random variables, $\vy^{(1)}, .., \vy^{(R)} \sim P_{\bx}(\vy ; A, L)$,
with the sample frequency $\vy$ represented as, $\vy = \frac{1}{R} \sum_{i=1}^R \vy^{(i)}$. We denote the \bcr problem with read sampling matrix $A$ by $\bcr(L,A)$. In its simplest form, $A$ can be constructed as follows,
\be
A_{ij} = \frac{\sum_{k=1}^{n_j-L+1} 1_{\{\lex^{-1}(i) = s_{j,k:k+L-1}\}}}{n_j-L+1}
\label{eq:MixMatrix}
\ee
This matrix represents {\it uniform} sampling of {\it error-free} reads along the sequence of the chosen species $j$,
assuming $L \leq n_j \: \forall j$. A non-zero element $A_{ij}$ means that read $i$ appears in the sequence of species $j$.
If $L > n_i$ we assume that the `tail' of each read is sampled uniformly from $\alphabet$ (see Appendix Section \ref{sec:appendix_small_seq}). %PROOFS

\begin{remark}
The above construction of $A$ assumes no read errors and no biases.
Incorporating more realistic sequencing models with non-uniform read density due to amplification biases, read errors (substitutions and indels), alignments errors etc. can be done by changing the definition of $A$ from eq. (\ref{eq:MixMatrix}). The same database $S$ may thus yield different matrices $A$, and the statistical and algorithmic properties of a certain \bcr problem depend on the database $S$ only through the matrix $A$.
The assumption in step $1$ is that species DNA fragments are sampled according to their DNA frequencies out of the total DNA present in a sample.
The model cannot accommodate deviations from this assumption which may arise from amplification biases and limited library complexity, which may distort the species frequencies - that is, the fraction of reads originating from a certain species may not represent the species' true frequency in the mixture. Accounting and correcting for such biases require analyzing multiple samples together.
\end{remark}

In similar to the read frequencies vector $\vy$, the matrix $A$ is
also huge ($4^L \times N$) but very sparse. In particular, the number of non-zero rows in $A$,
denoted $K$, is much smaller than $4^L$, as most of the rows in $A$ are zero and need not be stored. In the simple model above, $K \leq \sum_{j=1}^N (n_j - L + 1)$, which is roughly equal to the database size in nucleotides. In more complicated models involving read error, $K$ will be larger, but still much smaller than $4^L$. The computational complexity of our algorithms depends on $K$ (see Section \ref{sec:algorithm}). %PROOFS

An estimator $\hatx$ of the frequency vector $\bx$ is simply a function from the set of all reads and database, to the $n$-dimensional simplex, $\hatx: \Delta_{4^L} \times \mathcal{S} \to \Delta_N, \: \hatx = \hatx(\by, S)$ (here $\mathcal{S}$ is the set of all possible sequences databases, i.e. the space of all ordered finite collections of strings over $\alphabet$). %PROOFS

We can solve the \bcr problem by finding an estimator $\hatx$ minimizing an empirical loss function. That is,
define $\haty = A \hatx$, the empirical reads distribution given the estimator $\hatx$. We would like to minimize the loss
$l(\haty, y)$, and define the following estimator, %PROOFS
\be
\hatx = argmin_{\bx \in \Delta_N} l(A \bx, \by)
\label{eq:optproblem}
\ee
A natural loss function is the Kullback-Leibler divergence $l_{KL}(\by, \hat{\by}) = D(P_{\by} || P_{\hat{\by}})$. This formulation is equivalent to maximizing the likelihood of the data $\by$, according to the probabilistic model in eq. (\ref{eq:mixture_model}). Maximizing the likelihood using the EM algorithm was proposed in \cite{kessner2013maximum} for a very similar likelihood formulation - this approach, however, is currently not scalable to a large number of species. We choose instead the $l_2$ loss $l_{2}(\by, \hat{\by}) = ||\by - \hat{\by}||_2$, mainly for computational considerations. The $l_2$ loss leads to a standard optimization problem and many off-the-shelf solvers can be used.

We expect real mixtures to be sparse, with only a few hundreds to a few thousands species present
(out of hundreds of thousands). it is therefore appealing to use a sparsity-promoting loss in the cost function in eq. (\ref{eq:optproblem}), for example by penalizing $l_0$ norm of $\vx$. This is especially important when the number of reads is limited, to avoid over-fitting of the solution to the randomly sampled reads. The $l_0$ norm is not convex, leading to an intractable computational problem. The most common remedy of replacing the $l_0$ norm by the convex
$l_1$ norm does not work in our problem since for probability vectors in the simplex $\vx \in \Delta_N$ the constraint $||\vx||_1=1$ trivially holds. Promoting sparsity for probability distributions in the simplex by convex relaxation was recently proposed \cite{pilanci2012recovery}, but the approach does not scale to our problem's size. Instead, we developed a scalable divide-and-conquer thresholding algorithm (see Section \ref{sec:algorithm}) which minimizes the $l_2$ error, while enforcing sparsity implicitly, by a repeated truncation of non-zero frequencies. The resulting solution is guaranteed to be sparse, while still keeping the $l_2$ error low as desired. %PROOFS

\vspace{-0.1cm}
\subsection{Evaluating the Solution: Metrics}
\vspace{-0.05cm}
To evaluate reconstruction accuracy, we need a measure comparing the reconstructed solution $\hatx$ with the correct solution $\bx$. %PROOFS
Different applications may require different metrics - for example, in some applications we may be interested only in the {\it identity}
of the species, while in other applications one would want to detect changes in {\it frequencies}. It may be important to identify
the particular species or strain, or one may be satisfied with coarser reconstruction at the genus or family level.
There are two major groups of performance metrics:

\begin{enumerate}
\item
Phylogenetically-Unaware criteria: These metrics take into account only the species identities and frequencies.
Examples include the $l_p$ norm between the two vectors, recall-precision and Jaccard index.
We use the simple $l_2$ norm as a representative of this group. This metric measures the deviation in species {\it frequencies} between the true and reconstructed solutions, $\metric_{l_2}(\vx, \hatx) = \sqrt{ \sum_{i=1}^M (x_i - \hat{x}_i)^2}$.

\item
Phylogenetically-Aware criteria: These metrics take into account the phylogenetic relationship between species.
The main intuition here is that identifying a species close to the true species is in fact almost as good as reconstructing
the correct species. Examples include unifrac \cite{lozupone2005unifrac}, weighted unifrac \cite{lozupone2007quantitative}, and DPCoA
\cite{pavoine2004dissimilarities}.
We propose a novel Phylogenetically-aware criterion, using a Mahalanobis distance, \\
$\metric_{MA}(\vx, \hatx; D) = \sqrt{ (\vx - \hatx)^{\top} D (\vx - \hatx) } = \sqrt{ \sum_{i,j} D_{ij} (x_i - \hat{x}_i) (x_j - \hat{x}_j) }$.
%\label{def:mahalanobis} % no space for equations

The matrix $D$ is constructed to capture the phylogenetic distance between species (for example, from the species \rRNA sequences
themselves). High (low) values of $D_{ij}$ correspond to pairs of
species $(i,j)$ which are closely-related (remote). For concreteness,
we choose specifically $D = A^{\top} A$, which represent the
similarity between species based on their \rRNA sequences. The
resulting Mahalanobis distance measures the agreement between the true and reconstructed solutions, in terms of both the species identities and their frequencies, while taking into account the similarities between closely related species.

\end{enumerate}

\vspace{-0.3cm}
\section{Species Identifiability}
\vspace{-0.1cm}
In this section we study species identifiability - that is, the ability to correctly identify the species and their frequencies as
the number of reads, $R$, goes to infinity.

\begin{definition}
We say that the problem $\bcr(L,A)$ is identifiable, if for every $\bx^{(1)} \neq \bx^{(2)} \in \Delta_N$, there exists $\by \in \Delta_{4^L}$ such that $P_{\bx^{(1)}}(\by; A; L) \neq P_{\bx^{(2)}}(\by; A; L)$.
\label{def:identifiability}
\end{definition}

Species identifiability captures fundamental limits of our ability to reconstruct the species frequency vector from the observed reads data.
If the problem $\bcr(L,A)$ is identifiable, then {\it in principle} it is possible to correctly reconstruct the species frequencies
vector $\bx$, since different vectors will generate different
distributions on the observed reads. If the problem is not identifiable, recovering the correct frequencies
vector $\bx$ may not be possible, regardless of the data size and computational resources available, since other (incorrect) frequency
vectors give rise to an identical distribution on the observed reads data.

The identifiability question is not unique to the \bcr problem, and arises more generally when reconstructing the identity of long sequences in a mixture using short reads. For example, % in a similar yet different context,
conditions for the identification of isoforms from RNA-seq data were given in \cite{hiller2009identifiability}. The different Isoforms in \cite{hiller2009identifiability} are analogous to the different species in our problem, yet the precise modeling assumptions and identifiability criteria are different in the two problems.
Identifiability is determined by both the similarity between the sequences of different species, and the read length.
Longer and more diverse sequenced regions provide more information on the DNA sequence of different species in the mixture, and allow to distinguish between the underlying species more easily. However, even when the sequenced regions are informative enough, short sequenced reads obtained from these region may map to multiple species, thus species identification can be hard when reads are too short.
We next formalize this intuition mathematically, showing how identifiability is determined by the input sequence database (and the read length $L$) through the matrix $A$, which represents the relation between the unknown vector $\vx$ and the observed data $\vy$ (see Appendix for
proofs of all Propositions),

\begin{proposition}
Let $\Aone$ be the matrix constructed from $A$, concatenated with an all $1$'s row vector ${\textbf 1_N}$,
$
\Aone \equiv \left(
  \begin{array}{c}
    A \\
    {\textbf 1_N} \\
  \end{array}
\right)
$. The reconstruction problem $\bcr(L,A)$ is identifiable if and only if $rank(\Aone) = N$.

\label{prop:identifiability}
\end{proposition}

As the read length increases, it becomes increasingly easier to distinguish between species,
\begin{proposition}
Assume $N > 4$. Suppose that the database $S$ is composed of $N$ distinct sequences such that no sequence is a substring of another sequence,
i.e. $s_{i,j:k} \neq s_{i'} \: \forall i \neq i' \in \{1,..,N\}, \forall j,k \in \{1,..,n_i\}$. Let $A^{(u,L)}$ be the sampling matrix obtained by {\it uniform} sampling of reads with read length $L$, according to eq. (\ref{eq:MixMatrix}).
Then there is a critical read length $L_c$, $1 < L_c \leq \max_i n_i$ such that the problem $\bcr(L,A^{(u,L)})$ is identifiable if and only if $L \geq L_c$.

\label{prop:identifiability_vs_L}
\end{proposition}

\begin{remark}
We assume that no sequence in the database is a substring of another database sequence for mathematical
convenience. This assumption usually holds in practice provided a long enough region is sequenced, and can be relaxed  while still obtaining similar identifiability results. In addition, we demonstrated identifiability for a uniform read sampling distribution, but a similar result can be obtained for other read sampling distributions. %PROOFS
\end{remark}

Species identifiability is a worst-case measure, as it requires {\it all} species to be identified correctly.
In practice, we may settle for a weaker notion - for example we would still consider a reconstruction as successful if all species except a small minority were identified correctly. We next define {\it partial} identifiability, which is a weaker property characterizing our ability to correctly reconstruct identities and frequencies of specific species, while for other species the reconstruction may remain ambiguous.

\begin{definition}
We say that the problem $\bcr(L,A)$ is partially identifiable for species $j$, if for any $\bx^{(1)}, \bx^{(2)} \in \Delta_N$
such that $P_{\bx^{(1)}} (\vy ; A, L) = P_{\bx^{(2)}} (\vy ; A, L) \: \forall \vy \in \Delta_{4^L}$, we have
$x^{(1)}_j = x^{(2)}_j$.
\label{def:partial_identifiability}
\end{definition}

We can check partial identifiability using the following proposition,
\begin{proposition}
The problem $\bcr(L,A)$ is partially identifiable for species $j$, if and only if the standard basis vector $\ve^{(j)} \in \Delta_N$ is orthogonal to the null-space of $\Aone$, that is $\Aone \vx = 0 \Rightarrow x_j=0 \: \forall \vx \in \mathbb{R}^N$. %PROOFS
\label{prop:partial_identifiability}
\end{proposition}

We present the identifiability properties achieved for real \rRNA data in the Appendix (Section \ref{sec:identifiability_rRNA_appendix}).

\vspace{-0.2cm}
\section{Reconstruction Error}
\vspace{-0.1cm}
\label{sec:upper_bounds}

While identifiability ensures that one can {\it in principle} reconstruct correctly
the species vector $\bx$, it essentially assumes an unlimited number of reads and computational power.
Here we study the reconstruction error in more realistic scenarios, with a finite number of reads.
We prove general rigorous upper-bounds on reconstruction error, in terms of the matrix $A$ and the number of reads $R$. In the Appendix (Section \ref{sec:simulation_results_appendix}) we examine the actual error achieved in practice using simulations. %PROOFS

The next proposition gives bounds on the approximation error of the true frequency vector $\bx^{*}$ by the estimator $\hatx$, which we obtain using the empirically-observed frequencies $\by$, %PROOFS

\begin{proposition}
Consider the problem $\bcr(L,A)$ with $R$ sequence reads, and let $\hatx$ be the estimator minimizing the $l_2$ loss,
$\hatx = argmin_{\bx \in \Delta_N} \loss_2(A \bx, \by)$. Then,

\begin{enumerate}
\item
Let $\lambda_{\min}(A^{\top}A)$ be the smallest eigenvalue of $A^{\top}A$. The Euclidian $l_2$ distance satisfies:
\vspace{-0.1cm}
\be
Pr\Big(\metric_{l_2}(\hatx, {\vx}^*) \leq \frac{2 + \sqrt{\log(1/\delta)}}{\sqrt{R\lambda_{\min}(A^{\top}A)} }\Big) \geq 1-\delta, \quad \forall \delta \in (0,1). %PROOFS
\label{eq:l2_bound}
\ee
\vspace{-0.1cm}

\item
The Mahalanobis distance with weight matrix $A^{\top} A$ satisfies:
\vspace{-0.1cm}
\be
Pr \Big( \metric_{MA}(\hatx, {\vx}^*; A^{\top} A) \leq \frac{2 + \sqrt{\log(1/\delta)}}{\sqrt{R}}\Big) \geq 1-\delta, \quad \forall \delta \in (0,1). %PROOFS
\label{eq:mahalanobis_bound}
\ee
\end{enumerate}

\label{prop:reconstruction_error_bound}
\end{proposition}
\vspace{-0.35cm}
The bound on the convergence rate of the $\metric_{l_2}$ error depends on spectral properties of the matrix $A^{\top} A$. This is related to the database coherence, or similarity between the sequences $S_i$, encoded as similarity between the rows of $A$. In particular, when the problem is non-identifiable, the matrix $A^{\top} A$ has a zero eigenvalue and the reconstruction error may be arbitrarily large.

In contrast, the Mahalanobis bound does not depend on the matrix $A$ or even the dimension $N$. Even if the problem is non-identifiable, we still achieve convergence under the Mahalanobis distance - yet the entries in the solution vector will not converge to the corresponding entries in the true frequencies vector $\vx$, i.e. the reconstruction may assign (part of) the abundance of a specific species to different, yet highly similar species.

\vspace{-0.2cm}
\section{Divide-and-Conquer Algorithm}
\label{sec:algorithm}
\vspace{-0.1cm}

Solving a large scale \bcr problem with hundreds of thousands of species is computationally challenging. Even computing and storing
the matrix $A$ is not trivial, let alone minimizing the loss $l(A\vx, \vy)$ in eq. (\ref{eq:optproblem}).
We developed a scalable divide-and-conquer thresholding approach to cope with large problems.
In a nutshell, the algorithm divides the species into distinct blocks, solves a reduced-size problem within each block, setting species with low frequency in the solution for each block to zero, merges solutions from different blocks and iterates to reduce problem size. For the reduced size sub-problems we minimize
the $l_2$ loss, resulting in a convex optimization problem in each block
which we solve (exactly) using the CVX convex optimization software package \cite{cvx,gb08}. We describe the algorithm in more details in the Appendix (Section \ref{sec:algorithm_appendix}). %PROOFS

We implemented the divide-and-conquer algorithm in the \pack Matlab package (with some computationally demanding parts implemented in C). For a problem of size $N \sim\!5 \times 10^5$, running time is a few hours on a standard PC. The algorithm showed accurate reconstruction performance on simulated and real sequence data (see Section \ref{sec:simulation_results_appendix} and \cite{amir2013}).

\vspace{-0.15cm}
\section{Discussion}
\vspace{-0.1cm}

We formulated the \bcr problem mathematically, proposed an algorithm for solving large scale problems,
and obtained results on reconstruction performance.

We applied our approach on the \rRNA gene. However, the approach is generic and could be applied to other genes or regions.
The reconstruction performance is determined by properties of the genomic region used (in our case, \rRNA\!\!). Different genes or regions
will provide different information allowing us to distinguish between different species or strains, for example using clade-specific markers \cite{segata2012metagenomic}.

Extending our method to genome-wide metagenomics sequencing is possible, although computationally challenging. Our approach relies on the presence of a database of reference sequences, and cannot be used as is for {\it de novo} discovery
of new species. Currently there are $\sim\!3000$ whole-genome sequences in the NCBI database \cite{haft2012high}, compared to $\sim\!10^6$ \rRNA sequences in the Greengenes database, thus the current utility of the whole-genome approach is limited,
although it can be useful as a first filter before the remaining reads can be used for {\it de novo} discovery (assembly). More importantly, as these database are likely to grow in the near future, it will become increasingly appealing to use whole-genome sequencing, especially for identifying small variations in very close strains, or newly born alleles in present strains (where the \rRNA sequences may be identical and not allow identification). %PROOFS

Providing efficient algorithms for the \bcr problem is important - solving the \bcr problem directly for $N$ in the order of hundreds of thousands is currently infeasible due to memory and time issues. We used a feasible divide-and-conquer approach to cope with this problem yet there is still room for algorithmic improvements, especially when coping with read errors, which increase the size of the matrix $A$. Designing faster algorithms for handling larger databases will become crucial in light of the expected growth of microbial databases, in terms of both the number species and the regions (including whole-genomes) covered. % PROOFS

% The bibtex filename
\vspace{-0.2cm}
\bibliography{bib_metagenomics_nextgen}

\newpage % Appendix excluded from page-count
\renewcommand{\thesection}{\arabic{section}}
\section*{Appendix}
\setcounter{section}{1}

\renewcommand{\thesection}{\Alph{section}}
\subsection{Dealing with Sequences Shorter than the Read Length}
\label{sec:appendix_small_seq}
In rare cases the read length $L$ might be larger than the sequence length $n_j$ for a particular species $j$. For completeness,
we adopt a convention of a read having it's first $n_i$ nucleotides matching the sequence, and the next $n_i-L$
nucleotides distributed uniformly in $\alphabet^{L-n_i}$. In this case eq. (\ref{eq:MixMatrix}) generalizes to,

\be
A_{ij} = \frac{4^{\min(0, n_j-L)} \sum_{k=1}^{\max(1, n_j-L+1)} 1_{\{\lex^{-1}(i)_{1:\min(n_j,L)} = s_{j,k:k+L-1}\}}}{\max(1, n_j-L+1)}
\ee

\nin where $\lex^{-1}(i)_{1:k}$ denotes the first $k$ nucleotides in the i-th read (in lexicographic ordering).
One can adopt different conventions for this case, for example obtaining a shorter read (of length $n_j$), or using a `joker' symbol
for the tail (i.e. for example when sequencing the molecule $`AACGCT'$ a read of length $10$ will be $`AACGCTNNNN'$). The choice of different
conventions does not change our result significantly - we chose the above for mathematical convenience.

\subsection{Proof of Proposition \ref{prop:identifiability}}
\begin{proof}
From eq. (\ref{eq:mixture_model}), we have $P_{\bx}(\ve^{(i)} ; A, L) = [A \vx]_i, \: \forall i=1,..,4^L$. Therefore
 identifiability holds if and only if $A \vx^{(1)} = A \vx^{(2)} \Rightarrow \vx^{(1)} = \vx^{(2)}, \: \forall \vx^{(1)}, \vx^{(2)} \in \Delta_N$.

The vector $\Aone \vx$ is of size $4^L+1$, obtained as a concatenation of $A \vx$ with one additional entry, $[\Aone \vx]_{4^L+1} = \sum_{j=1}^N x_j$. For any $\vx \in \Delta_N$ the last entry $[\Aone \vx]_{4^L+1}$ is equal to $1$. Therefore $\Aone \vx^{(1)} = \Aone \vx^{(2)} \iff A \vx^{(1)} = A \vx^{(2)}, \: \forall \vx^{(1)}, \vx^{(2)} \in \Delta_N$.

If $rank(\Aone) = N$, we have $\Aone \vx^{(1)} = \Aone \vx^{(2)} \Rightarrow \vx^{(1)} = \vx^{(2)}, \: \forall \vx^{(1)}, \vx^{(2)} \in \mathbb{R}^N$. Therefore in particular the relation is true for any $\vx^{(1)}, \vx^{(2)} \in \Delta_N \subset \mathbb{R}^N$ and identifiability holds.

Conversely, if $rank(\Aone) < N$ then there exists a non-zero vector $\vx \in \mathbb{R}^N, \vx \neq {\textbf 0_N}$ in the null-space of $\Aone$. Thus $\Aone \vx = 0$ and in particular $[\Aone \vx]_{4^L+1} = \sum_{j=1}^N x_j = 0$.
Take a vector $\vx^{(1)} \in int(\Delta_N)$. Then there exists $\eps > 0$ such that $\vx^{(2)} \equiv \vx^{(1)} + \eps \vx \in \Delta_N$. But $A \vx^{(1)} = A \vx^{(2)}$ and $\vx^{(1)} \neq \vx^{(2)}$, therefore the problem $\bcr(L,S,A)$ is not identifiable.

\end{proof}
\qed % for some reason proof doesn't show qed square sign

\subsection{Proof of Proposition \ref{prop:identifiability_vs_L}}
\begin{proof}
Take $L=1$. Then the vector $\vy$ simply measures the fraction of `A's, `C's, `G's and `T's in the sample, and is of length $4$.
The matrix $A^{(u,1)}$ is of size $4 \times N$, and $rank(A^{(u,1)}) \leq 4$. Therefore, there exists a non-zero vector $\vx$ in the null-space of $A^{(u,L)}$, $A^{(u,L)} \vx = \vzero$. Let $\vx^{(1)} \in int(\Delta_N)$. %PROOFS
Then there exists $\eps > 0$ such that $\vx^{(2)} \equiv \vx^{(1)} + \vx \in \Delta_N$. But $P_x(\vx^{(1)}) = P_x(\vx^{(2)})$ for
$\vx^{(1)}, \vx^{(2)} \in \Delta_N$. Hence the problem $\bcr(1,A^{(u,1)})$ is not identifiable.

Take $L = n_{MAX} (=\max_i n_i)$. For each species $j$ define the read $r^{(j)} \equiv [S_j : {'A'}^{(L-n_j)}]$ where ${'A'}^{(k)}$
is a string of $k$ consecutive $'A'$s, and $[a:b]$ denotes the concatenation of the two strings $a$ and $b$. The read $r^{(j)}$ contains the sequence $S_j$, followed by a string of `A's. Since $S_{j'}$ is not a subsequence of $S_j$ for any $j \neq j'$, the read $r^{(j)}$ cannot appear when sequencing any other sequence $j' \neq j$, so $A_{\lex(r^{(j)}) j'} = 0 \: \forall j' \neq j$, and the $lex(r^{(j)})$-th row of $A$ is all zeros except for the j-th term.
This means that $A$ has $N$ independent rows, indexed by $\lex(r^{(1)}), .., \lex(r^{(N)})$ and $rank(A) = N$.
Therefore $rank(\Aone)=N$ and the problem $\bcr(n_{MAX},A^{(u,n_{MAX})})$ is identifiable.

Suppose that the problem is $\bcr(L,A^{(u,L)})$ is identifiable, and let $L' > L$. By definition, for every $\bx^{(1)} \neq \bx^{(2)} \in \Delta_N$,
there exists $\by \in \Delta_{4^L}$ such that $P_{\bx^{(1)}}(\by; A; L) \neq P_{\bx^{(2)}}(\by; A; L)$.
But the distribution $P_{\bx^{(i)}}(\cdot ; A; L)$ is obtained by a projection of the distribution $P_{\bx^{(i)}}(\cdot; A; L')$ (for $i=1,2$),
with $P_{\bx^{(i)}}(\cdot ; A; L) = \sum_{y', y = y'_{1:L}} P_{\bx^{(i)}}(\cdot; A; L')$.
Therefore, there must exist $\by \in \Delta_{4^{L'}}$ with $P_{\bx^{(1)}}(\by'; A; L') \neq P_{\bx^{(2)}}(\by'; A; L')$
and the problem $\bcr(L',A^{(u,L')})$ is also identifiable for $L'$.

\end{proof}
\qed % for some reason proof doesn't show qed square sign

\subsection{Proof of Proposition \ref{prop:partial_identifiability}}
\begin{proof}
In similar to Proposition \ref{prop:identifiability}, since $P_{\bx}(\ve^{(i)} ; A, L) = [A \vx]_i, \: \forall i=1,..,4^L$ we have
partial identifiability if and only if $A \vx^{(1)} = A \vx^{(2)} \Rightarrow \vx^{(1)}_j = \vx^{(2)}_j, \: \forall \vx^{(1)}, \vx^{(2)} \in \Delta_N$, which holds if and only if $\Aone \vx^{(1)} = \Aone \vx^{(2)} \Rightarrow \vx^{(1)}_j = \vx^{(2)}_j, \: \forall \vx^{(1)}, \vx^{(2)} \in \Delta_N$.

Assume that $\Aone \vx = 0 \Rightarrow x_j=0 \: \forall \vx \in \mathbb{R}^N$. Then, for any two vectors $\vx^{(1)}, \vx^{(2)} \in \Delta_N$
take $\vx = \vx^{(1)} - \vx^{(2)}$ to get,
\be
\Aone \vx^{(1)} = \Aone \vx^{(2)} \Rightarrow \Aone(\vx^{(1)} - \vx^{(2)}) = 0 \Rightarrow [\vx^{(1)} - \vx^{(2)}]_j=0 \Rightarrow x^{(1)}_j = x^{(2)}_j.
\ee

Therefore, $\bcr(L,A)$ is partially identifiable for species $j$.
For the other direction, assume that $\bcr(L,A)$ is partially identifiable for species $j$. Let $\vx \in \mathbb{R}^N$. Take
some $\vx^{(1)} \in int(\Delta_N)$ and set $\vx^{(2)} = \vx^{(1)} + \alpha \vx$ with $\alpha>0$ small enough such that
$\vx^{(2)} \in \Delta_N$. Then,

\be
\Aone \vx = 0 \Rightarrow \Aone \vx^{(1)} = \Aone \vx^{(2)} = 0 \Rightarrow x^{(1)}_j = x^{(2)}_j \Rightarrow x_j=0.
\ee

\end{proof}
\qed

\subsection{Identifiability in the \rRNA Database}
\label{sec:identifiability_rRNA_appendix}
\begin{figure}[H]
\begin{center}
\includegraphics[totalheight=0.35\textheight,width=0.7\textwidth]{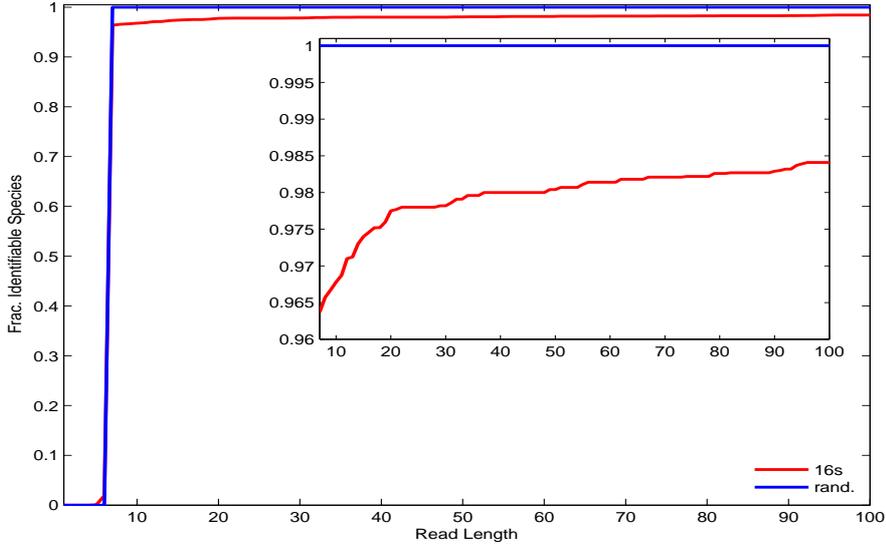} % pdf: 0.7, 0.8
\end{center}
\vspace{-0.05cm}
\caption[Species Identifiability in the Greengenes Database]{
Partial identifiability as a function of the read length. The red line shows results for a set of $N=10,000$ similar species from the Greengenes database. For comparison, the blue line shows results for $N=10,000$ sequences of the same length, with uniformly drawn i.i.d. characters. (i.e.
$Pr('A')=Pr('C')=Pr('G')=Pr('T)=0.25$ for each base).
The X-axis is read length used. The y-axis shows the fraction of identifiable species.
At $L=7$ we see a big jump in identifiability, as expected, since this is the point at which the number of equations $4^L$ exceeds the number of species $N$. For random sequences the problem is identifiable for $L\geq7$ (i.e., $100\%$ of species are partially identifiable). For the sequences from the \rRNA database, the vast majority ($\sim\!96.5\%$) of species are partially identifiable for $L=7$. The number of partially identifiable species then increases slowly with read length (see inset). Even at $L=100$ the problem is still not identifiable, but $\sim\!98.5\%$ of species can be identified. The remaining un-identified species contain groups of species with very close sequences, which can be distinguished only by increasing read length even further. %PROOFS
\label{fig:identifiability}}
\end{figure}

We checked the ability to identify species based on their \rRNA sequences. We downloaded the \rRNA Greengenes database from \url{greengenes.lbl.gov} \cite{desantis2006greengenes} (file `current\_prokMSA\_unaligned.fasta.gz', version dated 2010). After clustering together species with identical \rRNA sequences, we were left with $N= 455,055$ unique sequences of the \rRNA gene, with mean sequence length $1401$ - we refer to these $N$ unique sequences as the species. %PROFFS
We assume that the entire \rRNA gene is available - this can be achieved for example by shot-gun or RNA sequencing
(In practice, the choice of primers used when performing targeted DNA sequencing may be restricted due to biochemical considerations. This will affect the region sequenced and therefore all aspects of the reconstruction performance including identifiability - see \cite{amir2013}). Although the sequences are all distinct when considering the entire \rRNA sequences, identifiability is not guaranteed since we only observe short reads covering possibly non-unique portions of the \rRNA gene, which may cause ambiguities. %PROOFS
We plot in Figure \ref{fig:identifiability} the number of uniquely identifiable species as a function of the read length \readlen.
Even for very short \readlen, we can identify most species,
since the short reads aggregate information from the entire \rRNA gene. However,
even when \readlen\: is long ($\readlen=100$), there is still a small subset of species which are not identifiable.

\vspace{-0.05cm}

\subsection{Proof of Proposition \ref{prop:reconstruction_error_bound}}

\begin{proof}
Eq. (\ref{eq:optproblem}) with a $l_2$ loss implies that $A\bx$ is the Euclidean projection of $\by$ on the convex set $A(\Delta_N) \equiv \{\bz: \exists \vx \in \Delta_N, \: \bz=A\bx \}$ (namely, it is the closest point to $\by$ in $A(\Delta_N)$). Similarly, $A\bx^{*}$ is the Euclidean projection of $\by^{*}$ on $A(\Delta_N)$. Since projections on convex sets can only reduce distances \cite{rockafellar1970convex}, we have,
\be
\norm{A\bx-A\bx^{*}}_2 = \norm{A\bx-\by^{*}}_2 \leq \norm{\by-\by^{*}}_2.
\ee

The left hand side above is equal to the Mahalanobis distance,
since

\be
\metric_{MA}(\vx, {\vx}^*; A^{\top} A) = \sqrt{(\bx-\bx^{*})^{\top}(A^{\top}A)(\bx-\bx^{*})} = \norm{A\bx-A\bx^{*}}_2.
\ee

Therefore we get
\be
\metric_{MA}(\vx, {\vx}^*; A^{\top} A) \leq \norm{\by-\by^{*}}_2.
\label{eq:metric_inequality}
\ee

Recall that $\vy = \frac{1}{R} \sum_{i=1}^R \vy^{(i)}$ where the $\vy^{(i)}$ are i.i.d. vectors with $E[\vy^{(i)}] = \vy^*$.
Using large-deviation bounds on vectors \cite{shawe2004kernel} we get,
% it can be shown that with probability at least $1-\delta$ over drawing the $R$ reads (for any $\delta$) we have
\be
Pr \Big( \norm{\by-\by^{*}}_2 \leq \frac{2}{\sqrt{R}} + \sqrt{\frac{\log(1/\delta)}{R}} \Big) \geq 1-\delta , \quad \forall 0 < \delta < 1
\label{eq:metric_large_deviation}
\ee

Combining eqs. (\ref{eq:metric_inequality},\ref{eq:metric_large_deviation}), we get part $2$ of the proposition.
%\be
%Pr \Big( \metric_{MA}(\vx, {\vx}^*; A^{\top} A) \leq C \frac{\log(1/\delta)}{R} \Big) \geq 1-\delta, \quad \forall 0 < \delta < 1
%\ee

To prove part $1$, we need to convert this result to a bound on the Euclidian distance between $\bx$ and $\bx^*$.
The conversion is performed by first writing an eigen-decomposition of $A^{\top} A$, $A^{\top} A = U \Lambda U^\top$
where $U$ is an orthogonal matrix and $\Lambda$ a diagonal matrix with the eigenvalues of $A^{\top} A$.
This gives,

\begin{align}
\metric_{MA}(\vx, {\vx}^*; A^{\top} A)^2 &= (\bx-\bx^{*})^{\top}(U \Lambda U^{\top})(\bx-\bx^{*}) \nonumber \\
		        &\geq || U^{\top} (\bx-\bx^{*}) ||_2^2 \lambda_{\min}(A^{\top}A) \nonumber \\
		        &= ||(\bx-\bx^{*}) ||_2^2 \lambda_{\min}(A^{\top}A) \nonumber \\
		        &= \metric_{l_2}(\vx, {\vx}^*)^2 \lambda_{\min}(A^{\top}A)
\end{align}

Dividing both sides by $\lambda_{\min}(A^{\top}A)$, taking the square root and substituting in eq. (\ref{eq:mahalanobis_bound}) gives immediately part $1$.

\end{proof}
\qed % for some reason proof doesn't show the qed square sign

\subsection{Details of Divide-and-Conqour Algorithm}
\label{sec:algorithm_appendix}

\begin{framed}
{\bf Box 1: Divide-and-Conquer Reconstruction Algorithm} \\
\indent {\bf Input: $S$ - Set of Sequences, $\by$ - read measurements, Probabilistic model} \\
\indent {\bf Output: $\bx$ - vector of species frequencies} \\
\indent {\bf Parameters: $B$ - block size. $\tau_B$ - frequency threshold for each block. $k_{B,j}$ - number of partitions into blocks in $j$-th iteration, $k_F$ - final number of species allowed} \\
\begin{enumerate}

\item
Partition to blocks:
Set $\vv$ as a binary vector with one entry per species.
If this is the first partitioning, set iteration number $j=1$.
Repeat $k_{B,j}$ times:
\begin{enumerate}
\item
Partition species randomly into non-overlapping blocks of size $B$.
\item In each block (B) compute the matrix $A^{(B)}$, (where $^{(B)}$ denotes the restriction of a vector or a matrix to a block $B$),
and solve (exactly) the convex optimization problem (using CVX),
\be
\min_{\vx^{(B)}} ||A^{(B)} \vx^{(B)} - \vy||_2 \: \: s.t., x^{(B)}_i \geq 0
\label{eq:problem_representation_one_block}
\ee

\item Collect all species with frequency above the threshold:
if $x^{(B)}_i \geq \tau_B$, set $v_i = 1$. Set $j=j+1$.

\item Collect all linearly dependent species:
For each $i$ which is non-identifiable in the block (i.e. species $i$ is orthogonal to the null space of $A^{(B)}$)
set $v_i = 1$.

\end{enumerate}

\item Collect results from blocks:
Keep only indices $i$ with $v_i=1$, i.e. species with high enough frequency in at least one block reconstruction.

\item Reduce problem size:
Keep only species $i$ with $v_i=1$. Set $V = \{i, v_i=1\}$ and set
$A = A^{(V)}$, $\bx = \bx^{(V)}$. If $|V| > k_F$, go back to step 1.

\item Solve for the last time the $l_2$ minimization problem for the reduced matrix,

\be
\min_{\vx^{(V)}} ||A^{(V)} \vx^{(V)} - \vy||_2 \: \: s.t., x^{(V)}_i \geq 0
\label{eq:problem_representation_one_block}
\ee

Normalize $\vx^{(V)}$ to sum to one, and output the normalized vector as the solution
\end{enumerate}
\end{framed}

\subsection{Simulation Results}
\label{sec:simulation_results_appendix}

\vspace{-0.5cm} % pdf use -2
\begin{figure}[H]
\begin{center}
\includegraphics[totalheight=0.35\textheight,width=0.7\textwidth]{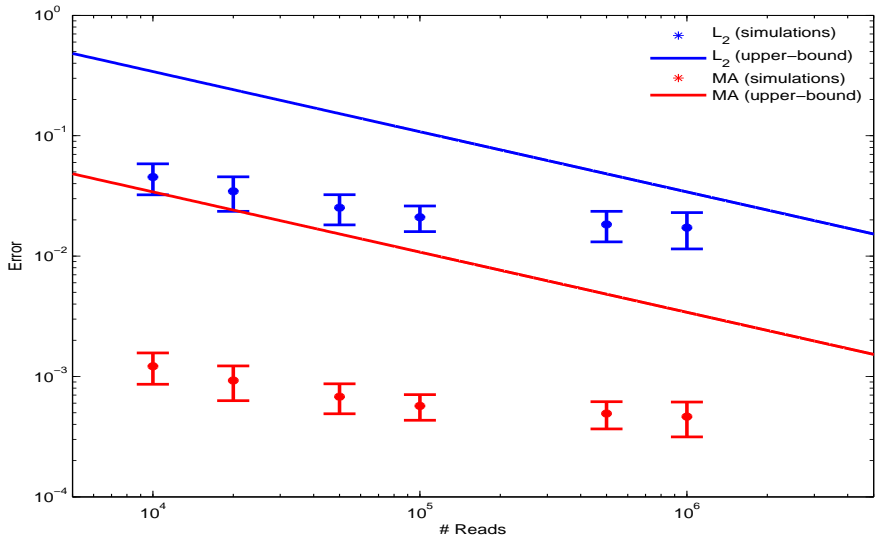} % pdf use 0.6, 0.8
\end{center}
\vspace{-0.5cm} % pdf use -2
\caption[Reconstruction Error as Function of Number of Reads]{
The curves show the $l_2$ (blue) and Mahalanobis (red) errors in
reconstruction for the example described in the text as function of
sample size (number of reads used). Error-bars show mean and 1 standard deviation of error over $100$ simulations. Solid curves show the theoretical upper-bounds, taken with $\delta=1/2$, giving a bound on the median error. For both metrics, the performance achieved in practice is significantly better than the upper bound.
\label{fig:reconstruction_error_vs_number_of_reads}}
\end{figure}

To evaluate the actual reconstruction performance in practice, we have performed a simulation study.
In Figure \ref{fig:reconstruction_error_vs_number_of_reads} we compare the actual reconstruction performance using simulations to the general rigorous bounds obtained in Section \ref{sec:upper_bounds}.

In our simulations, we studied the performance as a function of the number of reads using the Greengenes \rRNA database, with $N= 455,055$ unique
\rRNA sequences.
In each simulation we sampled at random $k=200$ species out of the total $N$.
We sampled the species frequencies from a power-law distribution with parameter $\alpha=1$, with frequencies normalized
to sum to one.
We then sampled sequence read according to the model in eq. (\ref{eq:mixture_model}). Read length was $\readlen = 100$.
The number of reads $R$ was varied from $10^4$ to $10^6$.

We performed reconstruction using Algorithm 1, with the following parameters:
block size $B=1000$, threshold frequency $\tau_B=10^{-3}$. The parameter $k_{B,j}$ represents a trade-off between time complexity and accuracy, and was initialized to $1$ at $j=1$, then set to $10$ when total size $|V|$ was below $150,000$.
Then, set to $20$ below $20,000$.  The final block size used was $k_F=1000$. %PROOFS

Very low error ($\sim\!2\%$) is achieved for $R > 500,000$,
showing that accurate reconstruction is possible for a feasible number of reads. The error rate achieved in practice is
much lower than the theoretical bounds, indicating that tighter bounds might be achieved.
There are many reasons for the gap between our bounds and simulation results: the concentration inequalities
we have used may not be tight, the particular frequency distribution chosen may perform better than the worst-case distribution, and most importantly, the small number of species present in the simulated mixture may enable accurate detection with a smaller sample size.
Proving improved bounds on reconstruction performance which consider all these issues including the sparsity of the solution is interesting yet challenging. Standard techniques (e.g. from compressed sensing) would need to be modified to achieve improved bounds since
they assume incoherence of the matrix $A$ which does not hold in our case, and do not consider the poisson sampling model we use for the reads.

\end{document}